\documentclass{article}

\IfFileExists{arxiv.sty}{\usepackage{arxiv}}{}

\usepackage[utf8]{inputenc}
\usepackage[T1]{fontenc}
\usepackage{microtype}
\usepackage{xcolor}

\usepackage{url}
\usepackage{graphicx}
\graphicspath{{fig/}}
\usepackage{booktabs}
\usepackage{amsfonts}
\usepackage{amsmath}
\usepackage{amssymb}
\usepackage{bm}
\usepackage{nicefrac}
\usepackage{siunitx}
\usepackage{float}

\usepackage{algorithm}
\usepackage{algorithmic}

\usepackage{tikz}
\usetikzlibrary{positioning,arrows.meta,fit}

\usepackage{caption}
\usepackage[hidelinks]{hyperref}
\usepackage[nameinlink,capitalise]{cleveref}

\usepackage[backend=biber,style=numeric,sorting=none]{biblatex}
\newcommand{\Dicke}[2]{\left|D^{#1}_{#2}\right\rangle}
\newcommand{\GO}{\mathrm{GO}}
\newcommand{\LO}{\mathrm{LO}}
\newcommand{\WAN}{\mathrm{WAN}}
\newcommand{\LAN}{\mathrm{LAN}}
\newcommand{\QLAN}{\mathrm{QLAN}}
\newcommand{\kreq}{k_{\mathrm{req}}}

\title{DH-EAC: Design of a Dynamic, Hierarchical Entanglement Access Control Protocol}

\author{
  Akihisa Takahashi \\
  College of Science and Engineering\\
  Aoyama Gakuin University\\
  \texttt{siva@rcl-aoayma.jp}
  \And
  Yoshito Tobe \\
  Department of Integrated Information Technology\\
  Aoyama Gakuin University\\
  \texttt{tobe@it.aoyama.ac.jp}
}

\IfFileExists{arxiv.sty}{\renewcommand{\shorttitle}{DH-EAC: Dynamic, Hierarchical EAC}}{}

\hypersetup{
  pdftitle={DH-EAC: Design of a Dynamic, Hierarchical Entanglement Access Control Protocol},
  pdfauthor={Akihisa Takahashi},
  pdfkeywords={Quantum networks, Entanglement access control, Dicke states, Fairness, Anonymity},
}

\begin{document}
\maketitle

\begin{abstract}
The quantum Internet aims to distribute entanglement among geographically separated quantum nodes, enabling applications such as quantum key distribution, distributed quantum computing, and cooperative quantum sensing. Since entanglement generation is probabilistic and quantum memory and concurrency are limited, efficient Entanglement Access Control (EAC) is a fundamental challenge. While a purely quantum Medium Access Control (MAC) scheme based on multipartite Dicke states has been proposed to resolve contention without classical round-trip communication, existing studies mainly focus on a single Quantum Local Area Network (QLAN). In wide-area quantum networks composed of multiple interconnected QLANs, it remains challenging to support hierarchical decision making, dynamic winner counts, and fairness under heterogeneous QLAN sizes.

This paper proposes Dynamic Hierarchical Entanglement Access Control (DH–EAC), a two-stage quantum lottery scheme for wide-area quantum networks. In the outer stage, a Global Orchestrator uses a Dicke state to uniformly select the winning QLANs and determine their allocation quotas. In the inner stage, each winning QLAN locally selects its winning nodes using another Dicke state. This hierarchical design resolves wide-area contention without introducing classical round-trip communication, while accommodating dynamic demands and mitigating size-induced bias. Simulation results show that, compared with a classical baseline, DH–EAC suppresses delay growth as the number of QLANs increases and maintains high node-level fairness even under skewed QLAN sizes.
\end{abstract}


\section{Introduction}
This section presents the background and motivation of this study, states the research goal, and summarizes the contributions and organization of the thesis.

\subsection{Background}
\label{sec:intro:background}
The quantum Internet aims to distribute quantum states over networks to enable applications such as distributed quantum computing and quantum key distribution \cite{Kimble2008,Wehner2018}.
Entanglement is a central resource, and the ability to generate and distribute entangled pairs and multipartite entangled states across quantum nodes directly affects application performance.
In practical quantum networks, however, (i) generation and distribution are probabilistic due to photon loss and decoherence, (ii) quantum memories have limited capacity and finite lifetime, and (iii) entanglement swapping at intermediate nodes is also probabilistic.
Therefore, when many requests arrive concurrently, a mechanism is required to decide who gets access to the limited entanglement generation and distribution resources, i.e., which nodes can access quantum resources and when.

In this thesis, we call the function that decides how entanglement resources are allocated to nodes (or node sets) Entanglement Access Control (EAC).
EAC plays a role analogous to Media Access Control (MAC) scheduling in classical networks, but quantum settings differ in two important ways: winner determination is inherently tied to measurement and state collapse, and post-lottery classical readjustment is not straightforward without consuming additional quantum resources.
Moreover, quantum-network operation often requires anonymity , fairness , and scalability.

Recently, a purely-quantum MAC based on Dicke states has been proposed, where the winner set is determined by local measurements only, eliminating classical round trips during the lottery loop \cite{Illiano2023EAC}.
While this design can potentially achieve anonymity and low latency by avoiding classical synchronization in the decision phase, existing studies mainly target a single quantum local-area network (QLAN).
In wide-area environments where multiple QLANs are connected over a wide-area network (WAN), additional challenges arise: hierarchical decisions across QLANs and within each QLAN, dynamically changing request sizes, and fairness under heterogeneous QLAN capacities.

\subsection{Research Goal}
\label{sec:intro:objective}
The goal of this study is to design an EAC protocol for wide-area quantum networks consisting of multiple QLANs that preserves the essential properties of purely-quantum lotteries while simultaneously satisfying anonymity, fairness, and scalability, and to establish an evaluation model for such protocols.
Specifically, we propose Dynamic Hierarchical Entanglement Access Control (DH--EAC), which realizes wide-area purely-quantum access control through a two-level lottery: an outer lottery for selecting winning QLANs and an inner lottery for selecting winning nodes within each winning QLAN.

\subsection{Contributions}
\label{sec:intro:contrib}
The contributions of this thesis are threefold.
First, we design a two-stage lottery that selects a winning QLAN set in the outer stage and winning nodes in the inner stage, and we present a protocol flow that avoids introducing classical RTTs (Round Trip Times) inside the lottery loop.
Second, we formalize a quota rule that determines how many winners (quota) each winning QLAN receives as a deterministic mapping conditioned on the outer-stage winner set, and instantiate it as an integer allocation under capacity constraints.
Third, we build a WAN/LAN-separated evaluation model that captures different loss and latency characteristics in the outer and inner stages, and we empirically study latency scaling, throughput, and node-level fairness (Jain's index) by comparing against classical baselines.

\subsection{Organization}
\label{sec:intro:structure}
Section~\ref{chap:background} introduces the background on quantum networks and EAC, and Section~\ref{chap:related} reviews related work.
Section~\ref{chap:problem} defines the system model and the problem statement.
Section~\ref{chap:method} describes the proposed DH--EAC protocol, and Section~\ref{chap:evaluation} presents the evaluation model and results.
Section~\ref{chap:discussion} discusses the implications and limitations, and Section~\ref{chap:conclusion} concludes the thesis and outlines future work.

\section{Preliminaries}
\label{chap:background}
This section summarizes background concepts on quantum networks and reviews Dicke states and coherent table lookup, which form the basis of the proposed protocol.

\subsection{Quantum Internet and Entanglement Distribution}
The quantum Internet provides functionalities that are not possible in classical networks by distributing qubits and entanglement over networks \cite{Kimble2008,Wehner2018}.
A typical stack consists of probabilistic entanglement generation over photonic links (link layer) and entanglement swapping to extend distance (network layer).
Because entanglement generation is probabilistic, failures require retries.
In addition, distribution must respect constraints such as the lifetime of quantum memories and maintaining high fidelity.

In this thesis, we view the network as a two-level hierarchy: WAN and local-area networks (LANs).
Specifically, we assume $m$ QLANs connected over a WAN, and nodes within each QLAN are connected over a LAN.
In many deployments, the WAN spans longer distances and therefore tends to suffer higher loss, while the LAN is shorter and has lower loss.
This difference directly affects both latency and success probability of EAC, and we explicitly separate WAN and LAN in the evaluation model (Section~\ref{chap:evaluation}).

\subsection{Entanglement Access Control}
Entanglement Access Control (EAC) decides which requests (or which node sets) are admitted given the entanglement generation and distribution capability available at a given time.
While similar roles are played by MAC and schedulers in classical networks, using classical round trips in the lottery loop may be undesirable in quantum settings because quantum states can decay during the wait and anonymity can be compromised.
Moreover, measurement is irreversible, and post-lottery re-selection or readjustment typically requires additional quantum resources.

For these reasons, a purely-quantum design that completes the lottery loop using quantum measurement only can match the unique requirements of quantum networks.

\subsection{Dicke States and Purely-Quantum Lotteries}
\label{sec:background:dicke}
A Dicke state is a symmetric multipartite state defined as the equal superposition of all computational-basis states of $n$ qubits with Hamming weight $k$:
\begin{equation}
\Dicke{n}{k}
=\frac{1}{\sqrt{\binom{n}{k}}}
\sum_{\substack{x\in\{0,1\}^{n}\\ |x|=k}}
|x\rangle .
\label{eq:dicke_def}
\end{equation}
If each qubit is distributed to a different participant and each participant performs a $Z$-basis measurement, exactly $k$ participants obtain outcome $1$ and the remaining participants obtain $0$, and the winning combination is uniform over all $\binom{n}{k}$ possibilities.
Thus, Dicke states can be used as a primitive for uniformly selecting $k$ winners out of $n$ participants without classical communication \cite{Illiano2023EAC}.

In our protocol, the outer stage uses $\Dicke{m}{K}$ to select $K$ winning QLANs uniformly from $m$ QLANs.
The inner stage uses $\Dicke{n_i^{\mathrm{avail}}}{k_i}$ to select $k_i$ winning nodes uniformly from the available nodes $n_i^{\mathrm{avail}}$ inside each winning QLAN~$i$.

\subsection{QROM and Coherent Decision Functions}
\label{sec:background:qrom}
A key idea of the proposed method is to design the protocol such that once the outer-stage winning QLAN set $S$ is determined by measurement, the quota vector $(k_1,\ldots,k_m)$ for that $S$ is also uniquely determined.
We explicitly define the mapping from a winning set $S$ to the quota as a deterministic decision function $g$ (Section~\ref{sec:method:quota}).

There are two main approaches to implementing $g$ as a quantum circuit.
For small instances, one can precompute a table and load it as a Quantum Read-Only Memory (QROM) \cite{Babbush2018PRXQROM}.
For larger instances, it can be more scalable to compute $g$ directly using reversible arithmetic.
A QROM primitive loads the data corresponding to an address register $|S\rangle$ into a data register, and design techniques exist to reduce T-count and depth of table lookup \cite{Babbush2018PRXQROM,Huggins2024Precompute}.
This thesis does not optimize QROM itself; instead, we make $g$ explicit at the protocol-specification level and assume that it can be implemented either by QROM or by reversible arithmetic.

\section{Related Work}
\label{chap:related}
This section reviews related work on entanglement distribution and resource allocation in quantum networks, purely-quantum medium access control and Entanglement Access Control (EAC), and circuit primitives relevant to coherent decision logic.

\subsection{Entanglement Distribution and Quantum-Network Simulators}
Entanglement distribution over quantum networks requires routing and link-layer protocols that cope with probabilistic generation and loss.
A recent survey summarizes routing and scheduling under constraints such as swap success probability, fidelity requirements, and limited quantum memory \cite{Abane2024Survey}.
Representative studies include efficient routing for remote entanglement generation \cite{Li2021npj,Zeng2024Routing}, asynchronous routing \cite{Asynch2024AQS}, opportunistic routing \cite{Opportunistic2022}, and online or learning-based scheduling \cite{ESDI2023,DQN2025}.
Evaluation is often performed using discrete-event simulators such as NetSquid \cite{Coopmans2021NetSquid}.
These works primarily focus on routing and physical/link-layer mechanisms, whereas this thesis focuses on access control (lottery-based admission) and its interaction with a wide-area hierarchy.

\subsection{Purely-Quantum MAC Based on Dicke States}
Illiano et al.\ proposed a purely-quantum medium access control that uses Dicke states to select winners solely by local measurement, eliminating classical RTTs during the lottery loop \cite{Illiano2023EAC}.
This approach can provide strong anonymity properties because participants learn only their own outcomes during the decision phase.
Our work builds on the same primitive but extends the setting to a wide-area hierarchy with multiple QLANs and explicitly incorporates dynamic request sizes and heterogeneous QLAN capacities.

\subsection{Dicke-State Generation and Symmetric-State Preparation}
The practicality of Dicke-state based protocols depends on how efficiently multipartite Dicke states and related symmetric states can be prepared.
Short-depth Dicke-state preparation circuits have been proposed \cite{BaertschiEidenbenz2022}, and further depth reductions using mid-circuit measurements and feed-forward have also been studied \cite{Yu2024DickePolylog}.
In addition, variational circuit design for preparing states in the symmetric subspace has been explored \cite{Bond2025SymmetricPRR}.
This thesis assumes that such state-preparation techniques are available and focuses on protocol design and evaluation at the EAC layer.

\subsection{Coherent Lookups and QROM}
Implementing classical decision logic coherently is a recurring requirement in quantum algorithms and protocols.
Quantum read-only memory (QROM) provides a primitive for table lookup with optimized resource trade-offs \cite{Babbush2018PRXQROM,Huggins2024Precompute}, including explicit constructions for low-$T$-count data lookup.
Relatedly, quantum random access memory (QRAM) has been proposed as a general architecture for coherent memory queries \cite{Giovannetti2008QRAM}.
Subsequent work studies practical compilation and optimization of coherent data access, including QROM circuit optimization \cite{Phalak2022QROM} and automated compilation of quantum memory or data-loading procedures \cite{Sinha2022QMem}.
In this thesis, we specify the quota rule as a deterministic mapping $S\mapsto k(S)$ and assume that it can be realized either via QROM or via reversible arithmetic, so that evaluation and implementation remain consistent.

\subsection{Fairness and Anonymity in Resource Allocation}
Fairness has long been studied in classical network resource allocation, and Jain's index is widely used to quantify how evenly resources are distributed \cite{Jain1984}.
Anonymity in quantum protocols is often analyzed in terms of what information is revealed before a decision is finalized.
By combining a Dicke-state lottery with a hierarchical design and a deterministic quota rule, this thesis aims to preserve the anonymity benefits of purely-quantum selection while improving scalability and fairness in a wide-area setting.
\section{System Model and Problem Statement}
\label{chap:problem}
This section defines the target hierarchical quantum-network model and clarifies the EAC problem considered in this thesis.

\subsection{Network Model}
\label{sec:problem:system}
We consider a hierarchical quantum network as shown in Fig.~\ref{fig:arch}.
There are $m$ QLANs connected over a WAN.
Each QLAN $i$ is controlled by a local orchestrator (LO; $\LO_i$).
Within each QLAN, multiple end nodes exist, and we denote by $n_i^{\mathrm{avail}}$ the upper bound on the number of nodes that can participate (and be allocated) at the time of a lottery, which can vary over time due to quantum-memory states and job availability.
Above the QLANs, a global orchestrator (GO
) accepts application requests and is responsible for outer-stage state preparation and distribution.

\begin{figure}[t]
  \centering
  \begin{tikzpicture}[
    scale=0.95, transform shape,
    box/.style={draw, rounded corners, align=center, minimum width=26mm, minimum height=9mm},
    smallbox/.style={draw, rounded corners, align=center, minimum width=14mm, minimum height=6mm, font=\scriptsize},
    linkq/.style={-Latex, thick},
    labelbox/.style={font=\small, fill=white, inner sep=1.2pt},
    ]
    \node[box] (go) {Global\\Orchestrator\\$\GO$};

    \node[box, below left=16mm and 22mm of go] (lo1) {Local\\Orchestrator\\$\LO_1$};
    \node[box, below=16mm of go] (lo2) {Local\\Orchestrator\\$\LO_2$};
    \node[box, below right=16mm and 22mm of go] (lo3) {Local\\Orchestrator\\$\LO_3$};

    \node[smallbox, below=9mm of lo1, xshift=-8mm] (n11) {Node};
    \node[smallbox, below=9mm of lo1, xshift=8mm]  (n12) {Node};
    \node[smallbox, below=9mm of lo2, xshift=-8mm] (n21) {Node};
    \node[smallbox, below=9mm of lo2, xshift=8mm]  (n22) {Node};
    \node[smallbox, below=9mm of lo3, xshift=-8mm] (n31) {Node};
    \node[smallbox, below=9mm of lo3, xshift=8mm]  (n32) {Node};

    \draw[linkq] (go) -- (lo1);
    \draw[linkq] (go) -- (lo2);
    \draw[linkq] (go) -- (lo3);
    \node[labelbox, below=2mm of go] {Outer stage (WAN)};

    \draw[linkq] (lo1) -- (n11);
    \draw[linkq] (lo1) -- (n12);
    \draw[linkq] (lo2) -- (n21);
    \draw[linkq] (lo2) -- (n22);
    \draw[linkq] (lo3) -- (n31);
    \draw[linkq] (lo3) -- (n32);
    \node[labelbox, below=3mm of lo2] {Inner stage (LAN)};
  \end{tikzpicture}
  \caption{Hierarchical quantum network considered in this thesis. The outer stage distributes quantum states from $\GO$ to each $\LO_i$ over the WAN, and the inner stage distributes states from each winning QLAN to nodes over the LAN. Classical messages used after the decision (e.g., notification and audit) are omitted for clarity.}
  \label{fig:arch}
\end{figure}

\subsection{Request Model}
An application sends a request $\mathrm{EAC\_Request}(\kreq,\mathrm{SLA})$ to $\GO$, including the required total number of winners $\kreq$ and an SLA (e.g., latency/fidelity thresholds and maximum retries).
$\GO$ collects only aggregated information $\{n_i^{\mathrm{avail}}\}_{i=1}^{m}$ from QLANs and does not handle individual node IDs.
The lottery consists of an outer stage (QLAN selection) and an inner stage (node selection), and the decision is finalized solely by local measurements.

\subsection{Outputs and Objective}
If a request succeeds, the protocol yields (i) the winning QLAN set $S\subseteq\{1,\ldots,m\}$, (ii) the quota $k_i$ assigned to each winning QLAN $i\in S$, and (iii) the set of winning nodes (within each QLAN).
Our objective is that the outer-stage measurement determines $S$ and $\{k_i\}$ simultaneously, and the inner-stage measurements determine the winning nodes, without requiring classical round trips to adjust the outcome.

\subsection{Anonymity and Threat Model}
In this thesis, \emph{anonymity} means that information that identifies winners is not unnecessarily shared before the lottery decision is finalized.
In purely-quantum MAC based on Dicke states, each node learns only its own outcome from its measurement and cannot directly learn the IDs of other winners during the decision phase \cite{Illiano2023EAC}.
Therefore, even within the same QLAN, it is not necessary to disclose the entire winner set to all participants during the lottery loop.

In DH--EAC, $\GO$ receives only aggregated values $\{n_i^{\mathrm{avail}}\}$ and does not store individual node IDs.
The outer-stage decision is made at the QLAN granularity, and only winning QLANs proceed to the inner-stage lottery.
In the inner stage, each node learns its own outcome by local measurement, making it difficult for non-winning nodes or other QLANs to identify winners in advance.
We assume that $\GO$ and each $\LO_i$ follow the protocol and focus on reducing information disclosure before the lottery is finalized.

In practice, some entity must identify the winning nodes to actually allocate entanglement resources.
We treat each $\LO_i$ as a trusted administrator within its QLAN and allow $\LO_i$ to confirm winners and perform allocation/audit \emph{after} the decision is finalized, consistent with \cite{Illiano2023EAC}.
We do not allow classical RTTs that change the lottery outcome, nor sending winner IDs before the decision is finalized.

\subsection{Fairness Metric}
We quantify fairness using Jain's fairness index for the node-level winning-probability vector $x=(x_1,\ldots,x_N)$:
\begin{equation}
J(x)=\frac{\left(\sum_{j=1}^{N}x_j\right)^2}{N\sum_{j=1}^{N}x_j^2}.
\label{eq:jain}
\end{equation}
$J=1$ indicates perfect fairness, and smaller values indicate higher bias toward a subset of nodes \cite{Jain1984}.
Our goal is to keep $J$ high even when QLAN sizes are heterogeneous.

\subsection{Notation}
Table~\ref{tab:symbols} summarizes the main symbols.

\begin{table}[H]
\centering
\caption{Main symbols}
\label{tab:symbols}
\begin{tabular}{ll}
\toprule
Symbol & Meaning\\
\midrule
$m$ & number of QLANs\\
$n_i^{\mathrm{avail}}$ & available nodes in QLAN $i$ at the lottery time (upper bound)\\
$\kreq$ & required total number of winners\\
$K$ & number of winning QLANs in the outer stage\\
$S$ & outer-stage winning QLAN set ($|S|=K$)\\
$k_i$ & quota assigned to QLAN $i$\\
$q^{\WAN},q^{\LAN}$ & per-attempt loss probability on WAN/LAN\\
$t_{\mathrm{dist}}^{\WAN},t_{\mathrm{dist}}^{\LAN}$ & distribution-time constants on WAN/LAN\\
$M$ & maximum number of retries\\
\bottomrule
\end{tabular}
\end{table}

\section{Proposed Method: DH--EAC}
\label{chap:method}
This section proposes Dynamic Hierarchical Entanglement Access Control (DH--EAC).
DH--EAC consists of a two-stage lottery---an outer stage across QLANs and an inner stage within each winning QLAN---and follows the design principle of avoiding classical RTTs inside the lottery loop.

\subsection{Overview}
\label{sec:method:overview}
DH--EAC first selects winning QLANs in the outer stage and then selects winning nodes in each winning QLAN in the inner stage.
In the outer stage, DH--EAC uniformly selects $K$ winning QLANs out of $m$ using the Dicke state $\Dicke{m}{K}$.
Once the outer-stage winning QLAN set $S$ is determined, a deterministic decision function $g$ uniquely maps $S$ (and the request/availability parameters) to a quota vector $k(S)=(k_1,\ldots,k_m)$, and the quota values are written coherently into auxiliary registers at each $\LO_i$.
In the inner stage, each winning QLAN $i\in S$ uniformly selects $k_i$ winning nodes out of the available nodes $n_i^{\mathrm{avail}}$ using the Dicke state $\Dicke{n_i^{\mathrm{avail}}}{k_i}$.
Both stages are finalized by measurement, so classical round trips for readjusting the outcome are not required.

\subsection{Outer Stage: QLAN Lottery and Quota Determination}
\label{sec:method:outer}
The outer stage consists of a preparation phase by $\GO$ and a measurement phase by each $\LO_i$.

\subsubsection{Selecting the Number of Winning QLANs $K$}
To satisfy a request of size $\kreq$, the total capacity of the winning QLAN set $S$ must satisfy $\sum_{i\in S} n_i^{\mathrm{avail}} \ge \kreq$.
Because the outer stage samples $K$ QLANs uniformly using $\Dicke{m}{K}$, $K$ must be chosen large enough so that resource shortage does not occur for any possible $S$.

We define the minimum value $K_{\min}$ as
\begin{equation}
K_{\min}=\min\left\{K:\ \sum_{j=1}^{K} n_{(j)}^{\mathrm{avail}} \ge \kreq \right\},
\end{equation}
where $n_{(j)}^{\mathrm{avail}}$ denotes the $j$-th smallest capacity after sorting in ascending order.
This ensures that even the most unfavorable combination (selecting $K$ QLANs among the smallest ones) can still satisfy the request.

In operation, one may add a safety margin $\beta$ to tolerate failures in state preparation/distribution.
However, the total quota must sum to $\kreq$, and if $K>\kreq$ then at least one winning QLAN must receive $k_i=0$, which is inconsistent with the interpretation that a winning QLAN participates in the inner-stage lottery.
Therefore, we impose $K\le \kreq$.
We adopt the following rule with a safety margin:
\begin{equation}
K=\min\!\left(\kreq,\ m,\ \left\lceil(1+\beta)K_{\min}\right\rceil\right).
\label{eq:safe_select_k}
\end{equation}
Algorithm~\ref{alg:safe_select_k} shows the pseudocode.

\begin{algorithm}[t]
\caption{SAFE--SELECT--K (Selecting $K$ with a safety margin)}
\label{alg:safe_select_k}
\begin{algorithmic}[1]
\REQUIRE $\kreq,\ \{n_i^{\mathrm{avail}}\}_{i=1}^{m}$, safety margin $\beta\in[0,1)$
\STATE Sort $\{n_i^{\mathrm{avail}}\}$ in ascending order; $s\gets 0$
\FOR{$K_{\min}=1$ \TO $m$}
  \STATE $s\gets s + n_{(K_{\min})}^{\mathrm{avail}}$
  \IF{$s \ge \kreq$} \STATE break \ENDIF
\ENDFOR
\IF{$s < \kreq$} \STATE \textbf{raise} \texttt{RESOURCE\_SHORTAGE} \ENDIF
\STATE $K \gets \min\!\left(\kreq,\ m,\ \left\lceil(1+\beta)K_{\min}\right\rceil\right)$
\STATE \textbf{return} $K$
\end{algorithmic}
\end{algorithm}

\subsubsection{Quota Decision Function $g$}
\label{sec:method:quota}
For a winning QLAN set $S$ with $|S|=K$, we define a mapping that uniquely determines the quota for each QLAN:
\begin{equation}
g:\ (S,\kreq,\{n_i^{\mathrm{avail}}\})\ \mapsto\ (k_1,\ldots,k_m).
\end{equation}
To avoid having protocol behavior depend on implementation details, we instantiate $g$ as an \textbf{integer allocation proportional to capacity} (largest remainder method).

We first assign a unit lower bound to each winning QLAN and forbid zero quota for winners:
\begin{equation}
k_i\ge 1\quad (i\in S),\qquad k_i=0\quad (i\notin S).
\end{equation}
We then distribute the remaining $r=\kreq-K$ slots proportionally to the residual capacity $\tilde{n}_i=n_i^{\mathrm{avail}}-1$ of each winning QLAN.
Specifically, we compute a fractional allocation
\begin{equation}
\hat{k}_i = 1 + r\cdot \frac{\tilde{n}_i}{\sum_{j\in S}\tilde{n}_j},
\end{equation}
and allocate the leftover slots in descending order of fractional parts (largest remainder method).
We also strictly enforce the capacity upper bound $k_i\le n_i^{\mathrm{avail}}$.
Algorithm~\ref{alg:quota_alloc} shows the pseudocode.

\begin{algorithm}[t]
\caption{QUOTA--ALLOC ($g$ with a unit lower bound)}
\label{alg:quota_alloc}
\begin{algorithmic}[1]
\REQUIRE $\kreq,\ \{n_i^{\mathrm{avail}}\},\ S$ ($|S|=K$)
\STATE \textbf{// 1) Reserve the unit lower bound}
\FOR{$i\in S$} \STATE $k_i\gets 1$ \ENDFOR
\FOR{$i\notin S$} \STATE $k_i\gets 0$ \ENDFOR
\STATE $r \gets \kreq - K$ \COMMENT{$K\le\kreq$ is required}
\IF{$r=0$} \STATE \textbf{return} $\{k_i\}$ \ENDIF
\STATE \textbf{// 2) Allocate the remaining slots proportionally to residual capacity}
\FOR{$i\in S$} \STATE $\tilde{n}_i \gets n_i^{\mathrm{avail}}-1$ \ENDFOR
\STATE $W \gets \sum_{i\in S}\tilde{n}_i$
\FOR{$i\in S$}
  \STATE $x_i \gets r\cdot \tilde{n}_i/W$
  \STATE $\Delta_i \gets \min\!\left(\lfloor x_i\rfloor,\ n_i^{\mathrm{avail}}-k_i\right)$
  \STATE $k_i \gets k_i + \Delta_i$
  \STATE $f_i \gets x_i - \lfloor x_i\rfloor$
\ENDFOR
\STATE $r' \gets \kreq - \sum_{i}k_i$
\WHILE{$r'>0$}
  \STATE $j \gets \arg\max_{i\in S:\ k_i<n_i^{\mathrm{avail}}} f_i$ \COMMENT{tie-breaking can be fixed or randomized}
  \STATE $k_j \gets k_j + 1$
  \STATE $f_j \gets 0$;\quad $r'\gets r'-1$
\ENDWHILE
\STATE \textbf{return} $\{k_i\}$
\end{algorithmic}
\end{algorithm}

\subsubsection{Coherent Allocation via $U_g$}
After defining $g$ as a classical function, we implement its reversible version as a unitary $U_g$.
Conceptually, when the outer-stage register $|Q\rangle$ represents $S$, we realize
\begin{equation}
U_g:\ |S\rangle|0\cdots 0\rangle \mapsto |S\rangle|k_1(S)\cdots k_m(S)\rangle
\end{equation}
so that the quota values are coherently written into the quota registers.
Each $\LO_i$ needs a local quota register large enough to represent $k_i\in[0,n_i^{\mathrm{avail}}]$, i.e.,
$\lceil\log_2(n_i^{\mathrm{avail}}+1)\rceil$ qubits.

For small instances, $U_g$ can be realized via QROM with $S$ as the address; for larger instances, it can be realized as reversible arithmetic that computes $g$ directly.
The key point is that $g$ is uniquely specified at the protocol level so that all $\LO_i$ derive the same quota by the same rule.

\subsection{Inner Stage: Node Lottery within Each Winning QLAN}
\label{sec:method:inner}
For each winning QLAN $i\in S$, after obtaining the quota $k_i$, $\LO_i$ selects $k_i$ winning nodes out of $n_i^{\mathrm{avail}}$ using a purely-quantum lottery.
Specifically, $\LO_i$ prepares $\Dicke{n_i^{\mathrm{avail}}}{k_i}$ and distributes the qubits to nodes; each node performs a $Z$ measurement, and the nodes with outcome $1$ are the winners.
This inner-stage lottery also does not require classical RTTs and is finalized by local measurements.

\subsection{Output Format: GHZ and EPR Resources}
After the Dicke-state lottery finalizes the set of winning nodes, the entanglement resource form required by an application can vary.
If a multipartite GHZ state is required, the state can be kept as is.
If EPR pairs (Bell pairs) are required, they can be obtained from GHZ via local operations and classical communication (LOCC).
Because such conversions do not change the lottery outcome, we treat them as part of the management-plane operations after the decision.

\subsection{Implementation Considerations}
\label{sec:method:impl}
Two major implementation considerations are (i) the circuit depth and fault tolerance for preparing Dicke states, and (ii) how to implement the quota unitary $U_g$ (QROM versus reversible arithmetic).
For the former, shallow Dicke-state preparation circuits have been proposed \cite{BaertschiEidenbenz2022,Yu2024DickePolylog}, suggesting that the multipartite states required for lotteries may be realizable within practical depth.
For the latter, QROM-based lookup is simple for small instances, while the table size grows as $\binom{m}{K}$; for larger instances, directly computing $g$ as reversible arithmetic can be more scalable.
This thesis focuses on protocol-layer design and the evaluation model, leaving detailed resource estimation and circuit-level optimization as future work.

\subsection{Overall Protocol Flow}
Finally, Algorithm~\ref{alg:overall} summarizes the overall DH--EAC protocol.
In the outer stage, $U_g$ coherently writes quotas, and each $\LO_i$ measures to determine win/loss and quota.
Only winning QLANs proceed to the inner-stage lottery.

\begin{algorithm}[H]
\caption{DH--EAC protocol (overall flow)}
\label{alg:overall}
\begin{algorithmic}[1]
\REQUIRE $\mathrm{EAC\_Request}(\kreq,\mathrm{SLA})$
\STATE $\GO$ collects $\{n_i^{\mathrm{avail}}\}$ and computes $K\gets\mathrm{SAFE\text{-}SELECT\text{-}K}(\kreq,\{n_i^{\mathrm{avail}}\})$
\STATE $\GO$ prepares the outer-stage state $\Dicke{m}{K}$ and quota registers, and coherently applies $U_g$
\STATE $\GO$ distributes the outer-stage (and quota) registers to each $\LO_i$
\STATE \textbf{The following steps can be executed in parallel across QLANs.}
\FOR{$i=1$ \TO $m$}
  \STATE Outer measurement $d_i\gets Z$-measurement
  \IF{$d_i=1$}
    \STATE \textbf{(QLAN $i$ wins)}
    \STATE Measure the quota register to obtain $k_i$
    \STATE $\LO_i$ prepares $\Dicke{n_i^{\mathrm{avail}}}{k_i}$ and distributes to nodes
    \STATE Each node performs a $Z$ measurement, finalizing the winner set $\mathrm{WIN\_SET}_i$
    \STATE $\LO_i$ sends an ACK to $\GO$ over the management plane (e.g., fidelity info, $\mathrm{hash}(\mathrm{WIN\_SET}_i)$)
  \ELSE
    \STATE Discard and terminate
  \ENDIF
\ENDFOR
\STATE $\GO$ aggregates ACKs and performs SLA checks and application notification
\end{algorithmic}
\end{algorithm}

This completes the description of DH--EAC. The next section evaluates the protocol.

\section{Evaluation}
\label{chap:evaluation}
This section evaluates the proposed DH--EAC using a numerical model and compares it against a strong hierarchical baseline that uses classical coordination in the outer stage.
We focus on (i) success probability under WAN/LAN losses, (ii) end-to-end latency and throughput scaling as the number of QLANs $m$ increases, and (iii) fairness under heterogeneous QLAN sizes.

\subsection{Evaluation Questions}
We study the following questions.
\begin{itemize}
\item \textbf{(Q1) WAN sensitivity.} How does DH--EAC's end-to-end success probability change as WAN loss increases, compared to a scheme that avoids WAN quantum distribution inside the lottery loop?
\item \textbf{(Q2) Throughput scaling.} Is there a realistic regime in which DH--EAC achieves higher throughput than a strong classical-coordination baseline, thanks to avoiding $m$-dependent control-plane costs?
\item \textbf{(Q3) Fairness.} Under skewed QLAN sizes, does the quota rule $g$ maintain high node-level fairness (Jain's index)?
\end{itemize}

\subsection{Schemes Compared}
\label{sec:eval:schemes_new}

\paragraph{DH--EAC (proposed).}
DH--EAC runs a Dicke-state lottery across QLANs in the outer stage, coherently writes per-QLAN quotas via $U_g$, and then runs a Dicke-state lottery inside each winning QLAN in the inner stage.

\paragraph{CH--EAC (baseline: classical outer stage + EAC inner stage).}
In this baseline, $\GO$ selects the winning QLAN set $S$ and computes quotas $k(S)$ \emph{classically} using the same rules as DH--EAC (SAFE--SELECT--K and QUOTA--ALLOC).
$\GO$ then broadcasts the decision $(S, \{k_i\})$ to all LOs over the classical control plane.
Only the winning QLANs run the inner-stage EAC lottery (Dicke state) to select $k_i$ nodes.

\subsection{Evaluation Model}
\label{sec:eval:model_new}
We use a WAN/LAN-separated model.
Per-attempt loss probabilities are denoted by $q^{\WAN}$ and $q^{\LAN}$, and each distribution is retried up to $M$ times.

\subsubsection{EAC-Consistent Success Probability}
A key point is that the original EAC analysis shows that, under a lossy distribution model with a retry budget $M$, the success probability of resolving contention for \emph{$k$ winners} is
\begin{equation}
P_{\mathrm{EAC}}(q,M;k)=\left(1-(q)^{M}\right)^{k},
\label{eq:succ_eac_k}
\end{equation}
which depends on the required number of winners $k$ rather than the total number of contenders \cite[Prop.~3]{Illiano2023EAC}.
We adopt Eq.~\eqref{eq:succ_eac_k} consistently for the \emph{inner stage} of both schemes.

\subsubsection{Success Probability $P$}
Let $p^{X}=1-(q^{X})^{M}$ for $X\in\{\WAN,\LAN\}$.

\paragraph{DH--EAC (Option A: protocol-consistent WAN distribution).}
In the outer stage, $\GO$ distributes the Dicke-register qubit to \emph{every} $\LO_i$ (because $\Dicke{m}{K}$ has $m$ participants) and also distributes the quota register as described in Algorithm~\ref{alg:overall}.
The quota register for QLAN $i$ requires
\begin{equation}
\ell_i=\left\lceil\log_2\!\left(n_i^{\mathrm{avail}}+1\right)\right\rceil
\end{equation}
qubits to represent $k_i\in[0,n_i^{\mathrm{avail}}]$.
Therefore, the total number of WAN-distributed qubits in the outer stage is
\begin{equation}
m+\sum_{i=1}^{m}\ell_i,
\end{equation}
and the outer-stage success probability is modeled as
\begin{equation}
P_{\mathrm{out}}^{\mathrm{D}}=\left(p^{\WAN}\right)^{m+\sum_{i=1}^{m}\ell_i}.
\end{equation}
In the inner stage, the total number of winners is $\sum_{i\in S}k_i=\kreq$, so by Eq.~\eqref{eq:succ_eac_k} we model
\begin{equation}
P_{\mathrm{in}}^{\mathrm{D}}=\left(p^{\LAN}\right)^{\kreq}.
\end{equation}
The end-to-end success probability is
\begin{equation}
P^{\mathrm{D}}=P_{\mathrm{out}}^{\mathrm{D}}\cdot P_{\mathrm{in}}^{\mathrm{D}}.
\label{eq:succ_dheac_new}
\end{equation}

\paragraph{CH--EAC.}
CH--EAC does not use WAN quantum distribution inside the lottery loop, so its success probability depends only on the inner stage:
\begin{equation}
P^{\mathrm{CH}}=\left(p^{\LAN}\right)^{\kreq}.
\label{eq:succ_cheac}
\end{equation}

\subsubsection{Latency $L$}
We model retry costs via the expected number of attempts for a truncated geometric process:
\begin{equation}
a(q,M)=\mathbb{E}[\min(T,M)]=\sum_{t=1}^{M}q^{t-1}=\frac{1-q^{M}}{1-q}.
\label{eq:expected_trials_new}
\end{equation}
We assume that different QLANs proceed in parallel, and max-type latencies capture straggler effects.

\paragraph{DH--EAC.}
In the outer stage, each $\LO_i$ must receive $1+\ell_i$ qubits over its WAN link, and the stage completes when the slowest LO completes:
\begin{equation}
L_{\mathrm{out}}^{\mathrm{D}}
\approx
t_{\mathrm{gen}}
+a(q^{\WAN},M)\,t_{\mathrm{dist}}^{\WAN}\cdot \max_{i\in\{1,\ldots,m\}}(1+\ell_i)
+t_{\mathrm{meas}}.
\end{equation}
In the inner stage, only winning QLANs participate, and the stage completes when the slowest winning QLAN completes:
\begin{equation}
L_{\mathrm{in}}^{\mathrm{D}}(S)
\approx
\max_{i\in S}\!\left(
t_{\mathrm{gen}}
+a(q^{\LAN},M)\,t_{\mathrm{dist}}^{\LAN}\cdot k_i(S)
+t_{\mathrm{meas}}
\right).
\end{equation}
The total latency is modeled as
\begin{equation}
L^{\mathrm{D}}\approx L_{\mathrm{out}}^{\mathrm{D}}+\mathbb{E}_{S}\!\left[L_{\mathrm{in}}^{\mathrm{D}}(S)\right].
\label{eq:lat_dheac_new}
\end{equation}

\paragraph{CH--EAC.}
CH--EAC replaces the quantum outer stage by classical coordination.
\begin{equation}
L_{\mathrm{ctrl}}^{\mathrm{CH}} \approx t_{\mathrm{ctl}} + t_{\mathrm{proc}}\,m + t_{\mathrm{stragg}}\log_2(m),
\label{eq:lat_ctrl}
\end{equation}
where $t_{\mathrm{ctl}}$ is the base WAN control RTT/broadcast latency, $t_{\mathrm{proc}}$ captures per-QLAN processing (e.g., authentication, policy evaluation, logging), and $t_{\mathrm{stragg}}\log_2(m)$ captures tail amplification due to stragglers.
The total latency is
\begin{equation}
L^{\mathrm{CH}}\approx L_{\mathrm{ctrl}}^{\mathrm{CH}}+\mathbb{E}_{S}\!\left[L_{\mathrm{in}}^{\mathrm{D}}(S)\right],
\label{eq:lat_cheac}
\end{equation}
where the inner stage is identical to DH--EAC (same $g$ and same $S$ distribution).

\subsubsection{Throughput and Fairness}
We define throughput as
\begin{equation}
\mathrm{THR}=\frac{P}{\mathbb{E}[L]}.
\end{equation}
Fairness is evaluated using Jain's index in Eq.~\eqref{eq:jain}.
Assuming a uniform distribution over outer-stage winning sets, the node-level winning probability can be computed as
\begin{equation}
\Pr[\text{node }u\in \QLAN i \text{ wins}]
=
\frac{1}{n_i^{\mathrm{avail}}}\,\mathbb{E}_{S}\!\left[k_i(S)\cdot \mathbf{1}[i\in S]\right].
\label{eq:node_win_prob_new}
\end{equation}

\subsection{Setup of Experiments}
\label{sec:eval:setup_new}

\subsubsection{Network Generation}
Given $m$ and a skew parameter $s$, we generate QLAN sizes using Zipf weights $w_i\propto 1/i^{s}$ and scale them so that the average QLAN size is $\bar{n}=16$ with a minimum size $n_{\min}=8$.
$s=0$ corresponds to a uniform setting, and larger $s$ yields stronger skew where a few QLANs dominate.

\subsubsection{Parameter Grid and Constants}
Table~\ref{tab:eval_grid_new} summarizes the grid used in the plots.
We restrict WAN loss to at most $q^{\WAN}=0.05$ to reflect a moderate-loss regime for short-haul or well-engineered links.

\begin{table}[H]
\centering
\caption{Evaluation Parameter Grid (main figures)}
\label{tab:eval_grid_new}
\begin{tabular}{ll}
\toprule
Item & Values\\
\midrule
Number of QLANs $m$ & $\{4,8,16,32\}$\\
WAN loss $q^{\WAN}$ & $\{0.01,0.02,0.03,0.05\}$\\
LAN loss $q^{\LAN}$ & $\{0.01\}$\\
Request ratio $\kreq/\sum_i n_i^{\mathrm{avail}}$ & $\{0.10,0.20,0.40,0.60\}$\\
Skew $s$ & $\{0.0,0.5,1.0,1.5,2.0\}$\\
Max retries $M$ & $3$\\
\bottomrule
\end{tabular}
\end{table}

Unless otherwise stated, we set
$t_{\mathrm{gen}}=\SI{2}{ms}$,
$t_{\mathrm{dist}}^{\WAN}=\SI{0.5}{ms}$,
$t_{\mathrm{dist}}^{\LAN}=\SI{0.05}{ms}$,
and $t_{\mathrm{meas}}=\SI{1}{ms}$.
For the CH--EAC control plane, we set
$t_{\mathrm{ctl}}=\SI{4}{ms}$,
$t_{\mathrm{proc}}=\SI{0.05}{ms}$,
and $t_{\mathrm{stragg}}=\SI{0.5}{ms}$.
The safety margin for SAFE--SELECT--K is $\beta=0.10$.

\subsection{Results}
\label{sec:eval:results_new}

\subsubsection{Success Probability vs.\ WAN Loss (Q1)}
Fig.~\ref{fig:perf_curves_new} (left) shows success probability as a function of $q^{\WAN}$ for $m=16$, $s=1.0$, and request ratio $0.40$.
Because CH--EAC avoids WAN quantum distribution in the lottery loop, it remains nearly constant in $q^{\WAN}$ (it depends only on the LAN factor).
DH--EAC decreases as $q^{\WAN}$ increases because the outer stage distributes the Dicke-register qubit and quota registers to all $m$ LOs (Eq.~\eqref{eq:succ_dheac_new}).

\subsubsection{Throughput vs.\ The Number of QLANs (Q2)}
\begin{figure}[H]
  \centering
  \includegraphics[width=0.49\linewidth]{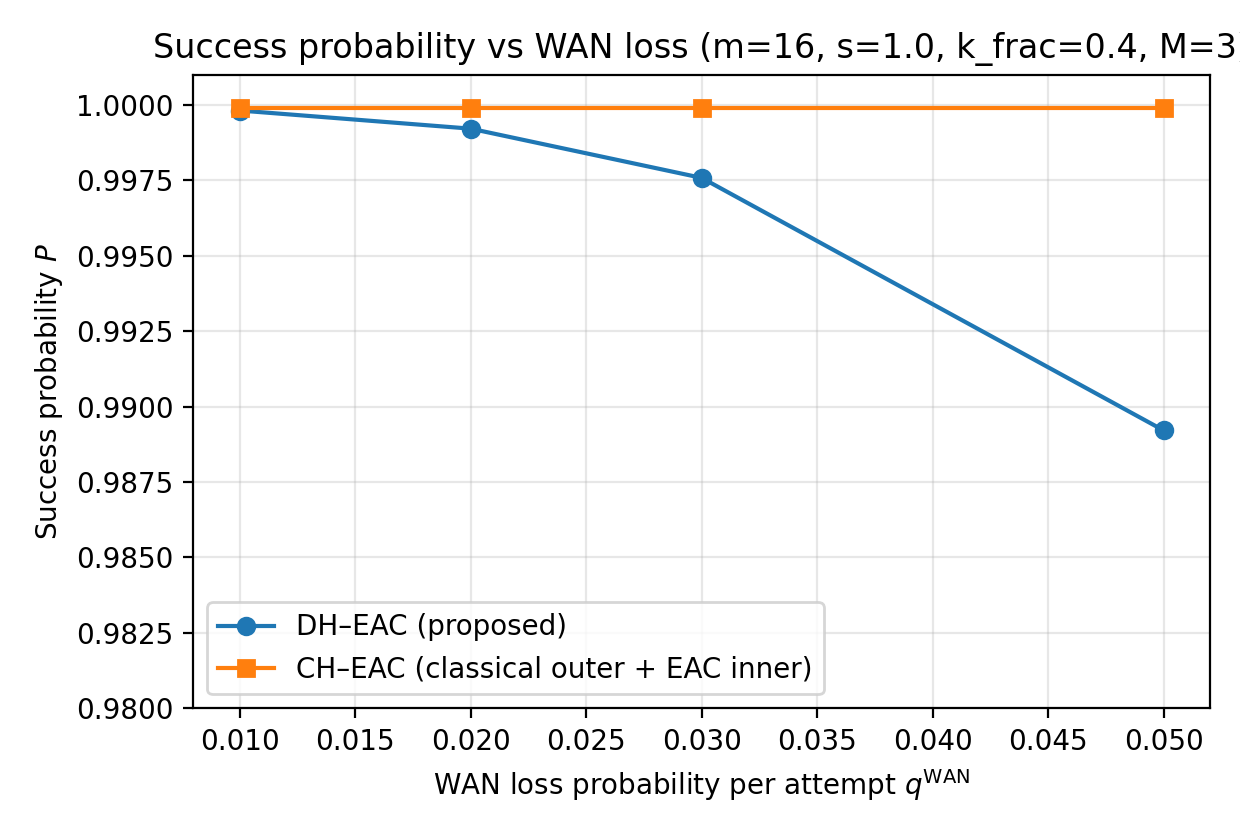}
  \includegraphics[width=0.49\linewidth]{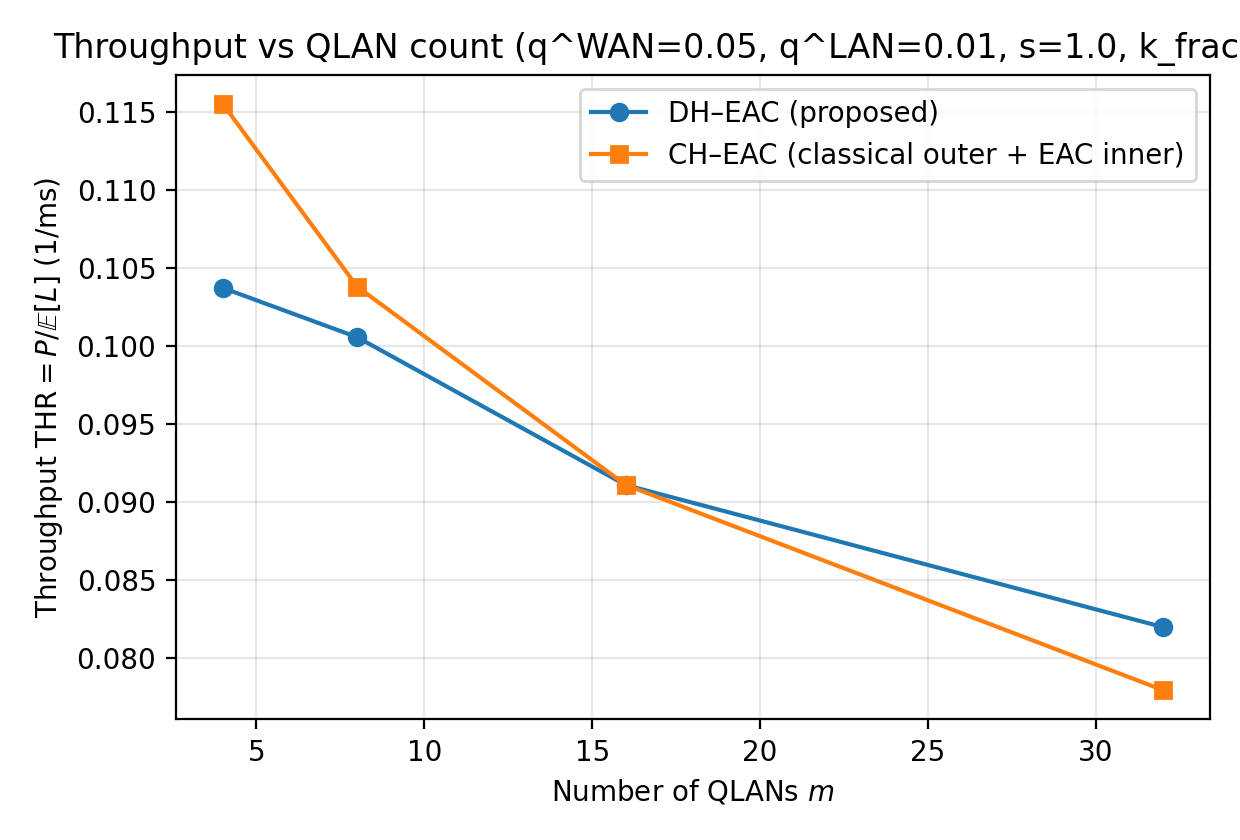}\\
  \caption{Performance curves (left: success probability vs.\ WAN loss; right: throughput vs.\ number of QLANs).}
  \label{fig:perf_curves_new}
\end{figure}

Fig.~\ref{fig:perf_curves_new} (right) plots throughput as $m$ increases under $q^{\WAN}=0.05$, $q^{\LAN}=0.01$, $s=1.0$, and request ratio $0.40$.
For small $m$, CH--EAC can have higher throughput because it pays no WAN quantum cost in the outer stage.
As $m$ grows, CH--EAC's classical coordination latency increases due to per-QLAN processing and straggler tails (Eq.~\eqref{eq:lat_ctrl}).
DH--EAC avoids classical RTTs inside the lottery loop and therefore exhibits better scaling, yielding a higher-throughput region at larger $m$.

\subsubsection{Break-Even Map (Q2)}
\begin{figure}[H]
  \centering
  \includegraphics[width=0.78\linewidth]{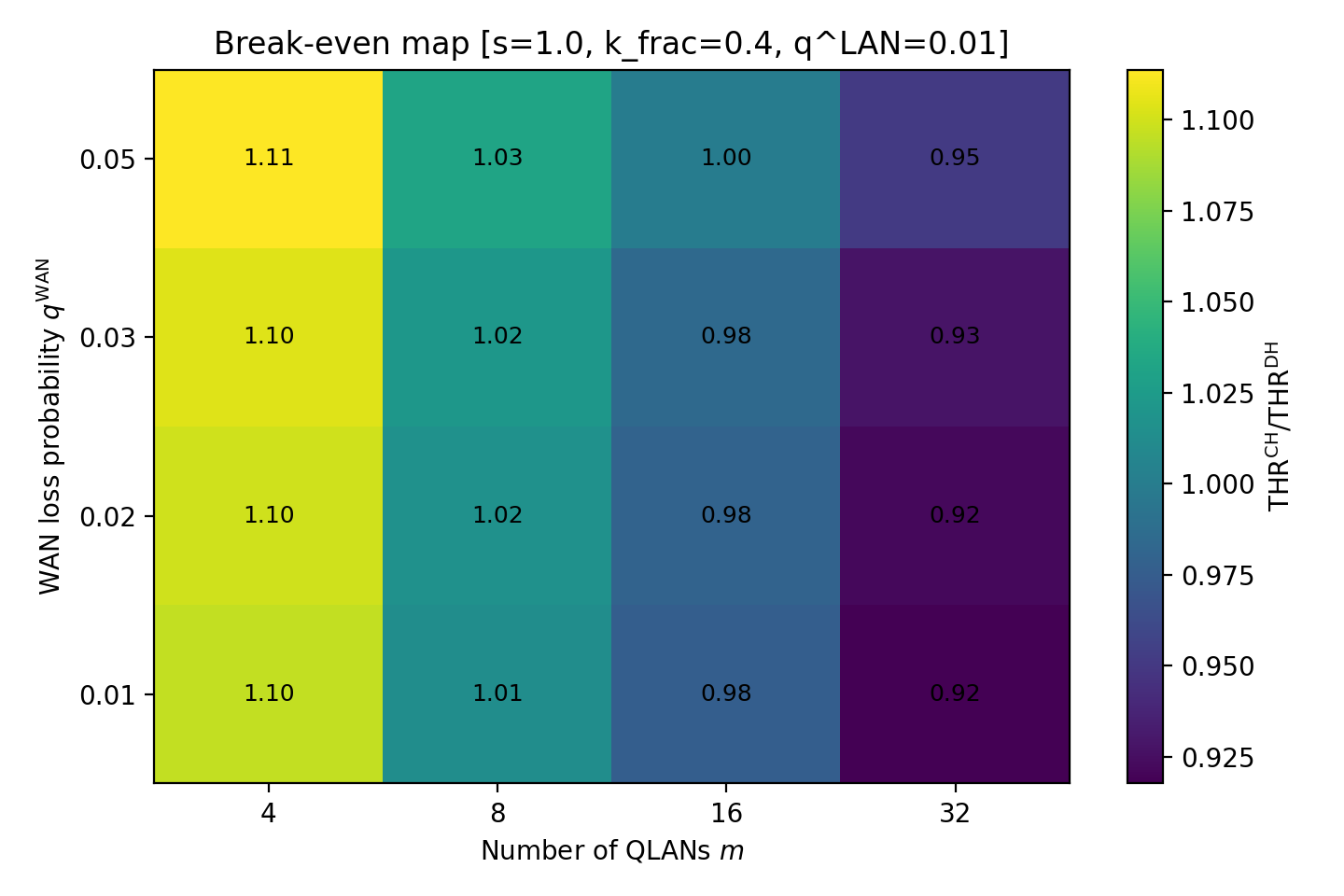}
  \caption{Break-even map against CH--EAC. Values $<1$ indicate that DH--EAC is superior in throughput.}
  \label{fig:breakeven_new}
\end{figure}

Fig.~\ref{fig:breakeven_new} shows the ratio $\mathrm{THR}^{\mathrm{CH}}/\mathrm{THR}^{\mathrm{D}}$ on the $(m,q^{\WAN})$ plane (fixed $q^{\LAN}=0.01$, $s=1.0$, request ratio $0.40$).
Values below 1 indicate that DH--EAC has higher throughput.
The map highlights a realistic \emph{break-even region}: DH--EAC tends to be advantageous when (i) $m$ is large enough for classical coordination overhead to dominate and (ii) WAN loss is moderate so that the outer-stage quantum distribution succeeds with high probability.

\subsubsection{Fairness (Q3)}
\begin{figure}[H]
  \centering
  \includegraphics[width=0.49\linewidth]{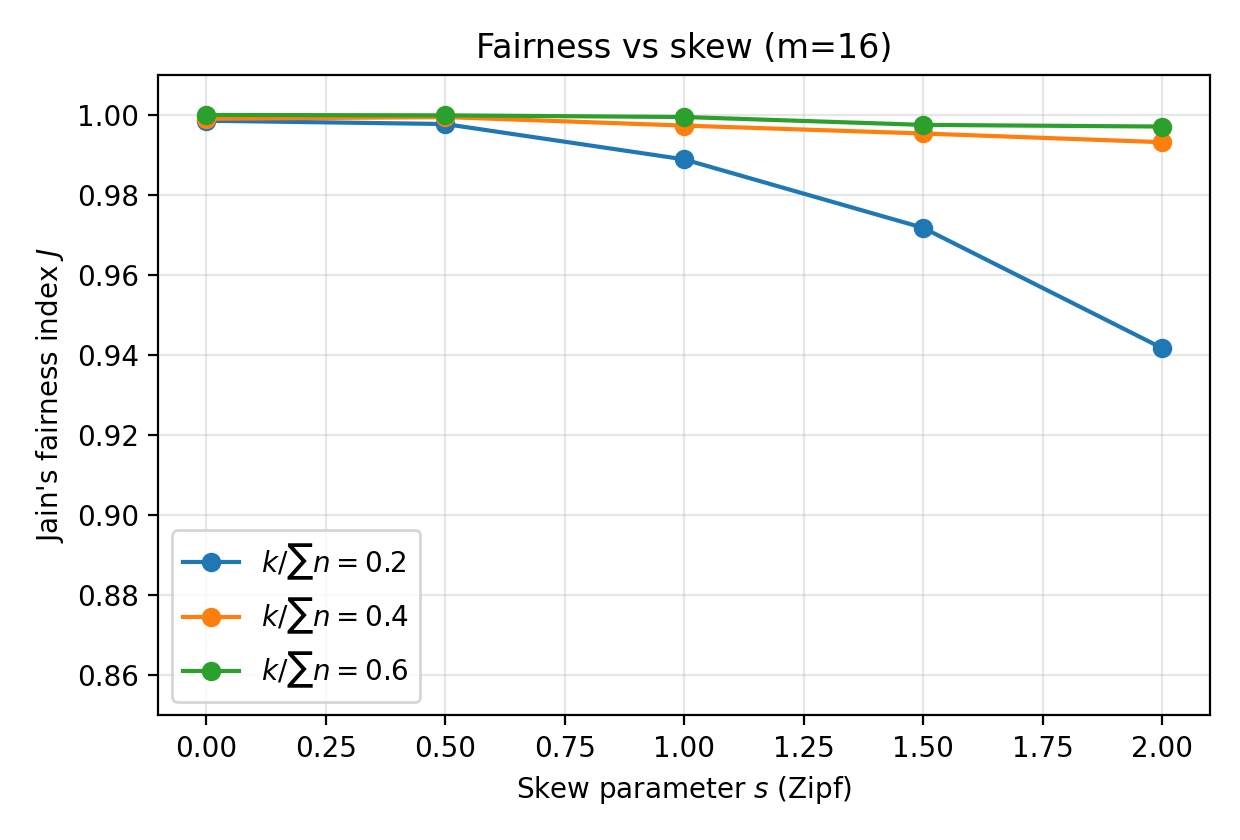}
  \includegraphics[width=0.49\linewidth]{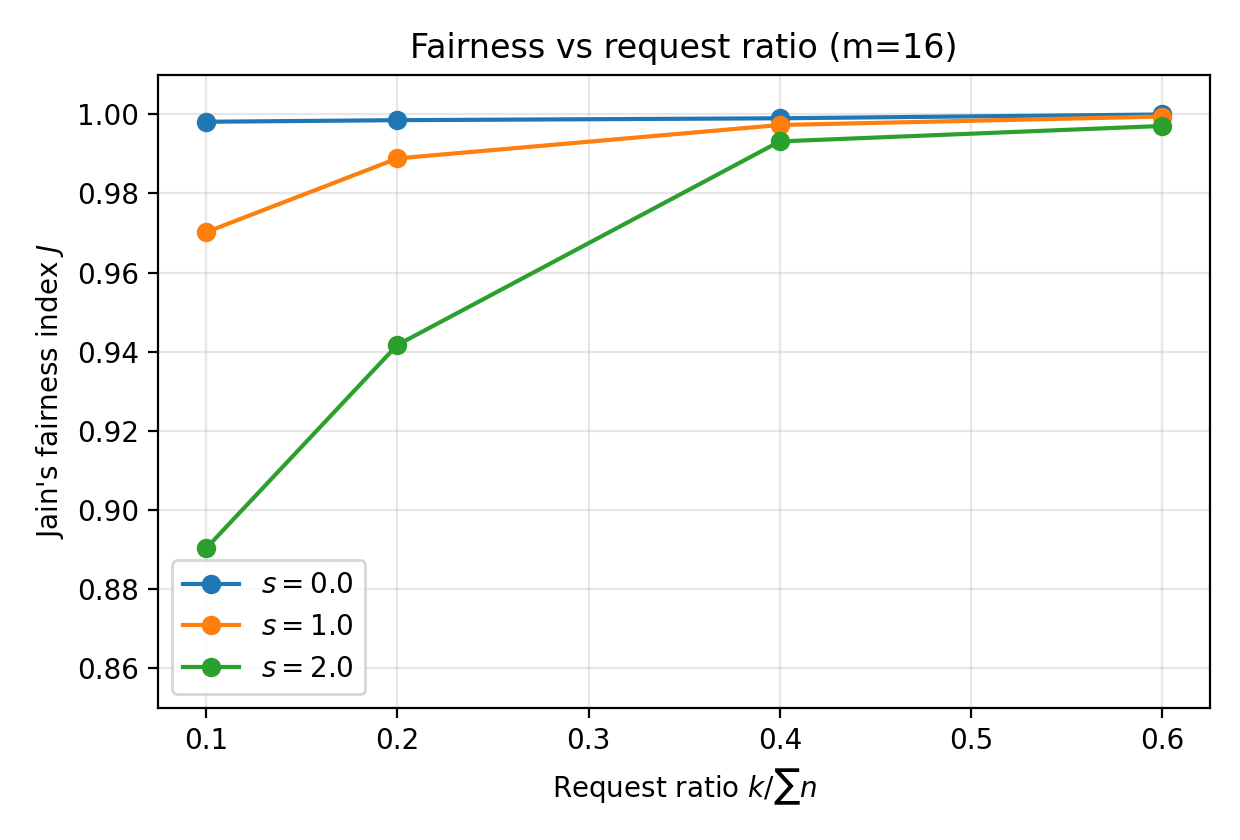}\\
  \caption{Fairness of DH--EAC (left: Jain's index vs.\ skew $s$; right: Jain's index vs.\ request ratio).}
  \label{fig:fairness1_new}
\end{figure}

Fig.~\ref{fig:fairness1_new} shows Jain's index trends for $m=16$.
As skew $s$ increases, QLAN sizes become more heterogeneous, but Jain's index remains close to 1 for medium-to-high request ratios.
Jain's index tends to increase as the request ratio increases because quotas and winners are distributed across more nodes.

\begin{figure}[H]
  \centering
  \includegraphics[width=0.49\linewidth]{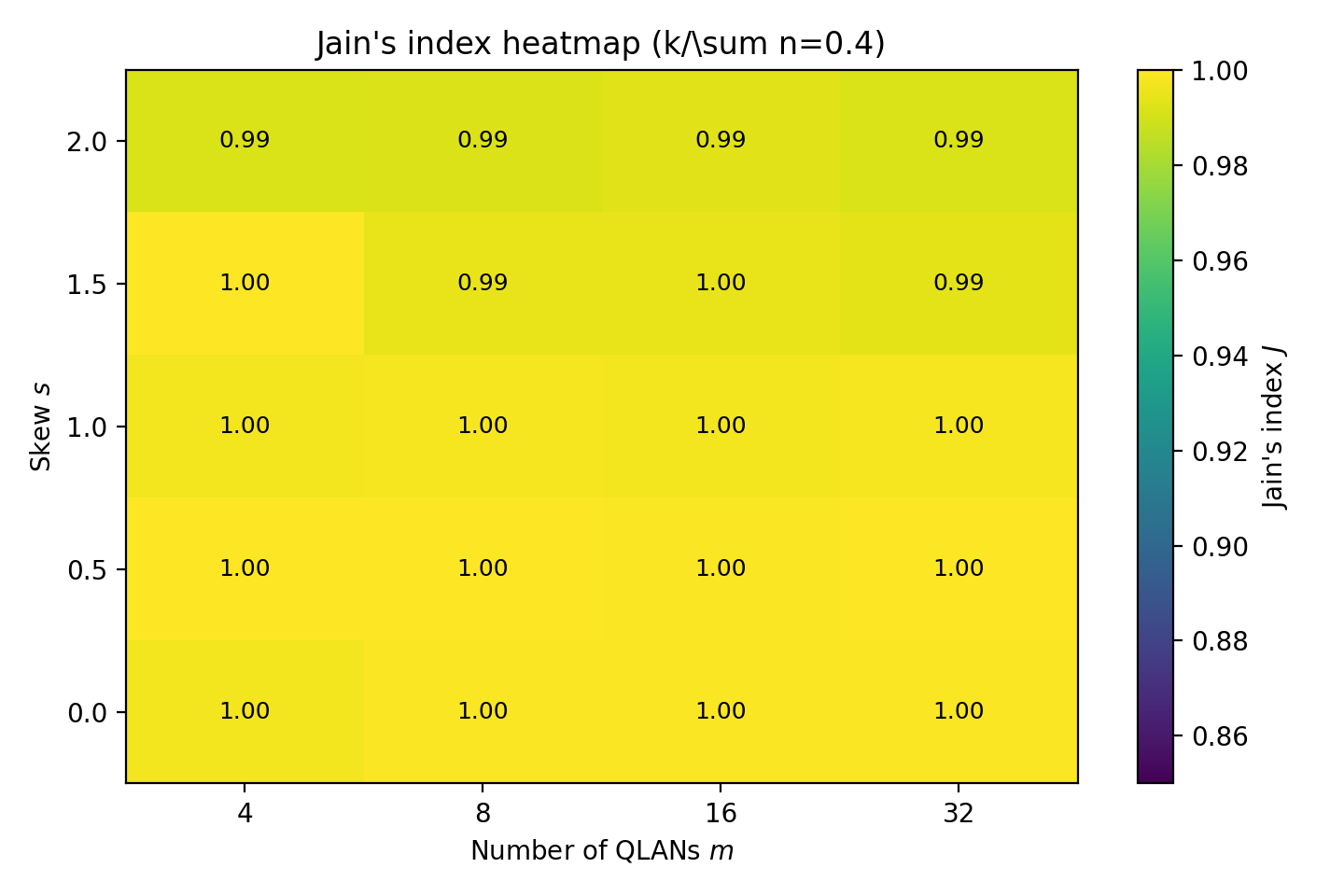}
  \includegraphics[width=0.49\linewidth]{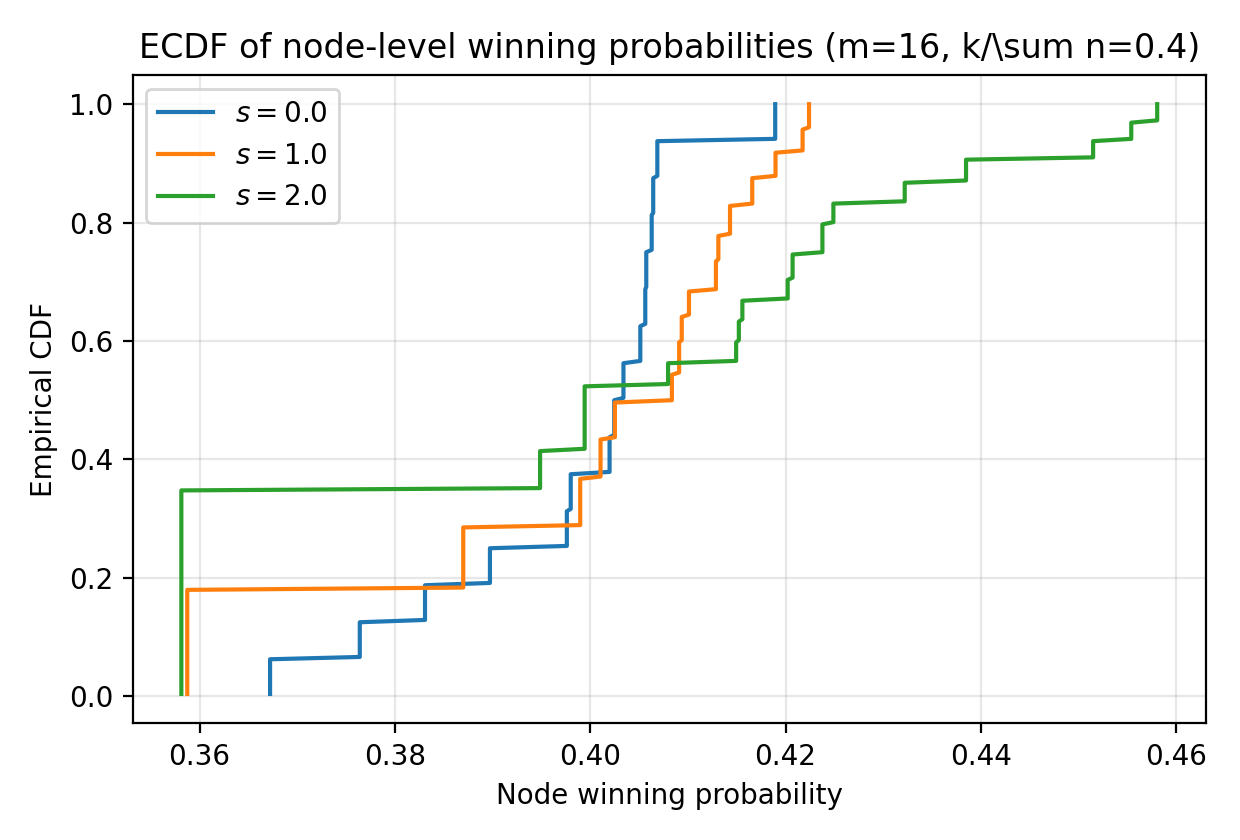}\\
  \caption{Distributional fairness evaluation (left: heatmap on $(m,s)$; right: ECDF of node-level winning probabilities).}
  \label{fig:fairness2_new}
\end{figure}

Fig.~\ref{fig:fairness2_new} shows a heatmap of Jain's index on the $(m,s)$ plane and the empirical CDF of node-level winning probabilities.
Across a wide range of parameters, DH--EAC maintains high fairness while allocating winners proportionally to capacity via $g$.

\subsection{Summary}
This section revised the evaluation model so that the outer-stage WAN quantum distribution is consistent with the DH--EAC protocol description: the Dicke-register qubit (and quota register) are distributed to all $m$ LOs, yielding an outer-stage factor $(p^{\WAN})^{m+\sum_i\ell_i}$.
DH--EAC can be disadvantaged in success probability when WAN loss increases because it performs WAN quantum distribution in the outer stage.
However, DH--EAC can achieve higher throughput in a realistic regime where the number of QLANs is large and WAN loss is moderate, because it avoids $m$-dependent classical coordination latency in the lottery loop.
We also observe that the quota decision function $g$ maintains high node-level fairness even under skewed QLAN sizes.

\section{Conclusion}
\label{chap:conclusion}
This thesis proposed DH--EAC, a quantum-native access-control protocol for wide-area quantum networks composed of multiple QLANs.
DH--EAC applies a two-stage lottery: an outer-stage Dicke-state lottery selects a set of winning QLANs, and inner-stage Dicke-state lotteries select winning nodes within each winning QLAN.
A central goal of the design is to keep the decision phase measurement-finalized, avoiding classical round trips that would adjust the outcome after the lottery has started.
To this end, we also specified a deterministic quota rule $g$ that maps the outer-stage winner set to a unique per-QLAN quota vector, allowing inner-stage lotteries to proceed without negotiation.

We developed an evaluation model that separates WAN and LAN effects and used it to study latency, throughput, and fairness.
The results indicate a clear trade-off.
Compared with classical hierarchical control, DH--EAC can incur additional WAN-related cost because it relies on WAN quantum distribution in the outer stage.
At the same time, by executing the outer stage once and parallelizing inner-stage lotteries across winning QLANs, DH--EAC avoids control-plane synchronization in the decision loop and can improve latency scaling as the number of QLANs increases.
We also observed that the quota decision rule used in this thesis tends to preserve high node-level fairness, even when QLAN capacities are heterogeneous.

Several directions remain for future work.
First, it is important to quantify circuit-level costs and fault-tolerance requirements for preparing Dicke states at scale and for implementing the reversible quota computation $U_g$, including a comparison between QROM-based lookup and direct reversible arithmetic.
Second, more realistic evaluation should incorporate physical-layer effects such as decoherence, fidelity degradation, and swapping failures, as well as interactions with routing and scheduling, for example by integrating DH--EAC into discrete-event simulators.
Finally, stronger operational and security mechanisms are needed for deployments beyond the honest-but-curious model, including audit and verification procedures in the management plane that preserve the principle that classical communication must not be used to alter the lottery outcome once it has started.

\printbibliography
\end{document}